# Quality-Based Conditional Processing in Multi-Biometrics: Application to Sensor Interoperability

Fernando Alonso-Fernandez, *Member, IEEE*, Julian Fierrez, *Member, IEEE*, Daniel Ramos, *Member, IEEE*, and Joaquin Gonzalez-Rodriguez, *Member, IEEE*

*Abstract*—As biometric technology is increasingly deployed, it will be common to replace parts of operational systems with newer designs. The cost and inconvenience of reacquiring enrolled users when a new vendor solution is incorporated makes this approach difficult and many applications will require to deal with information from different sources regularly. These *interoperability* problems can dramatically affect the performance of biometric systems and thus, they need to be overcome. Here, we describe and evaluate the ATVS-UAM fusion approach submitted to the *quality-based evaluation* of the 2007 BioSecure Multimodal Evaluation Campaign, whose aim was to compare fusion algorithms when biometric signals were generated using several biometric devices in mismatched conditions. Quality measures from the raw biometric data are available to allow system adjustment to changing quality conditions due to device changes. This system adjustment is referred to as quality-based conditional processing. The proposed fusion approach is based on linear logistic regression, in which fused scores tend to be log-likelihood-ratios. This allows the easy and efficient combination of matching scores from different devices assuming low dependence among modalities. In our system, quality information is used to switch between different system modules depending on the data source (the sensor in our case) and to reject channels with low quality data during the fusion. We compare our fusion approach to a set of rule-based fusion schemes over normalized scores. Results show that the proposed approach outperforms all the rule-based fusion schemes. We also show that with the quality-based channel rejection scheme, an overall improvement of 25% in the equal error rate is obtained.

## I. INTRODUCTION

### A. Biometric Systems Interoperability

THE increasing interest in biometrics is related to the number of applications where a correct assessment of identity is crucial [1]. Biometrics is used in many governmental and

Manuscript received December 1, 2008; revised July 9, 2009. Date of publication May 27, 2010; date of current version October 15, 2010. This work has been supported by BioSecure NoE, and the TEC2006-13141-C03-03 and TEC2006-13170-C02-01 projects of the Spanish Ministry of Science and Technology. The work of J. Fierrez was supported by a Marie Curie Fellowship from the European Commission. This paper was recommended by Guest Editor K. W. Bowyer.

The authors are with the ATVS/Biometric Recognition Group, Escuela Politecnica Superior, Universidad Autonoma de Madrid, 28049 Madrid, Spain (e-mail: fernando.alonso@uam.es; julian.fierrez@uam.es; daniel.ramos@uam.es; joaquin.rodriguez@uam.es).

Color versions of one or more of the figures in this paper are available online at http://ieeexplore.ieee.org.

Digital Object Identifier 10.1109/TSMCA.2010.2047498

civilian applications, offering greater convenience and advantages over traditional security methods based on something that you *know* (password, PIN) or something that you *have* (card, key, etc.). But using a single trait for recognition is affected by problems like noisy data, non-universality, lack of distinctiveness, spoof attacks, etc. [2]. Additional problems may arise when a biometric device is replaced (maybe from a different vendor) without reacquiring the corresponding template [3] or when templates generated with different proprietary algorithms are matched [4]. These *interoperability* problems, typically not overcome by biometric systems, affects the recognition performance, sometimes dramatically [5], [6]. Unfortunately, as biometrics is extensively deployed, it will be common to replace parts of operational systems as they are damaged or newer designs appear, or to exchange information among applications developed by different vendors or biometric data acquired in heterogeneous environments [3].

Multibiometric systems consolidate identity evidence from multiple sources, which helps alleviate many of these limitations because the different sources usually compensate for the inherent limitations of the others [2]. Integration at the matching score level is the most common approach (e.g., [7]) because it has the advantage of needing only the output matching scores of the different systems, which greatly facilitates the integration of multimodal biometric across different vendors (where the internal functionality of the system is often not disclosed). In this context, a measure of trust or quality of the data that defines the reliability of the recognition process can provide additional improvement, which helps optimize a structure lacking homogeneity while ensuring system interoperability by integrating data of different nature [8].

With this motivation, the *quality-based evaluation* of the BioSecure Multimodal Evaluation Campaign-BMEC [9] was organized with the aim of comparing fusion algorithms when biometric signals come from several devices in mismatched conditions. This paper describes the ATVS-UAM fusion strategy submitted to this evaluation [10], with outstanding results [second out of 13 system in terms of half total error rate (HTER) and fourth in terms of equal error rate (EER)] [9]. Face still samples collected with two cameras of different resolution and fingerprint samples collected with an optical and a thermal sensor were used. Quality information was also provided with the aim of adapting the fusion algorithms to the different devices. Data was extracted from the Biosecure Multimodal

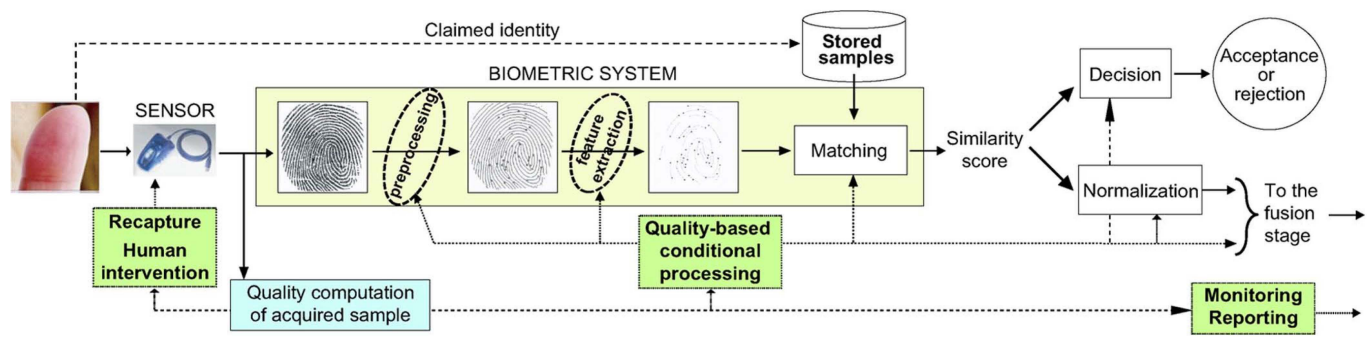

Fig. 1. Roles of a sample quality measure in the context of biometric systems.

Database [11], collected by 11 European institutions of the BioSecure Network of Excellence [12] between November 2006 and June 2007. This new database includes features not present in existing ones: more than 600 individuals acquired simultaneously in three different scenarios (over the Internet, in office environment with a desktop PC, and in indoor/outdoor environments with mobile devices) over two acquisition sessions, with different sensors for certain modalities.

### B. Related Work in Quality-Based Conditional Processing

Biometric quality assessment is an active field of research [13], with many quality assessment algorithms proposed such as [14]–[18]. Recent efforts have also been focused on the standardization of biometric quality information and its incorporation to biometric data structures [19].

In biometric systems working in verification mode, several steps are typically performed once a signal has been acquired (see Fig. 1 [1]): *1) preprocessing*, in which the input signal is enhanced to simplify subsequent steps; *2) feature extraction*, in which we further process the signal to generate a discriminative and compact representation; *3) matching*, where the feature representation of the input biometric signal is compared against the template corresponding to the claimed identity that is stored in the system database, resulting in a *similarity* or *matching score*; and *4) decision*, where the score is compared to a *decision threshold* in order to accept or reject the input identity claim. In multibiometric systems working at the matching score level, the output score is further combined with scores from other systems in a *fusion* stage to generate a new matching score that is then used for recognition. Prior to the fusion, the scores can be transformed to a common domain through a *normalization* step [20].

There are several roles regarding a quality measure in the context of biometric systems [19], [21], as shown in Fig. 1: *1)* monitoring tool [22] in order to accumulate statistics of the system (e.g., to identify sources experiencing problems); *2)* to recapture a sample not having enough quality; and *3)* to switch between different processing blocks of the system (*quality-based conditional processing* [19]). Since the work presented here falls in the last category, only related work in this domain will be covered next.

- **Preprocessing and feature extraction**: If the quality of the sample is low, we can invoke special enhancement algorithms. Also, we can use features robust to the kind of degradation that the biometric signal is suffering [23]. In some cases, there will be useless parts (e.g., damaged fingerprint regions) that can be discarded. The extracted features can be ranked depending on the quality of local regions. These information can be exploited afterward during the matching, as done, for example, in [24] and [25].

- **Matching and decision**: Depending on the quality of the acquired templates, we can use different matching algorithms (which also depend on the kind of features extracted) or adjust the sensitivity of the matcher to the quality of the signals under comparison [23]. Other works (e.g., [24] and [25]) give more weight to high quality features in the computation of the matching score.

- **Fusion**: Quality information has been incorporated in a number of fusion approaches; for instance, weighting results from the multiple sources depending on the quality [8] or dynamically switching them [25]. Instead of using a weighting scheme, the method in [27] estimates the joint densities of the matching score and the quality of the genuine and impostor classes. The work in [28] exploits the dependence between matching scores and quality measures by introducing the latter as an additional dimension of the classification problem. Finally, a novel device-specific quality-dependent score normalization technique is presented in [29], which is used for matching samples coming from different devices.

### C. Our Proposed Approach and Contributions

Our approach for the BMEC rely on the use of linear logistic regression fusion [30], [31], a trained classification fusion method which works in a probabilistic framework. Two hypotheses are defined for each comparison: *target* (the compared biometric data comes from the same individual) and *nontarget* hypothesis (the compared data comes from different individuals). Prior to fusion, the score of each single modality is mapped to a log-likelihood-ratio among the target and nontarget hypotheses, according to a Bayesian framework [32]. This mapping process is known as *calibration*, and has been proposed for speaker recognition [33]. It is achieved by *linear logistic regression*, which has been recently used for calibration purposes in speaker recognition [30], [31]. This allows the efficient combination of scores originating from different biometric devices, as is the case of the quality-based evaluation. If all sources are independent and independence assumptions and a

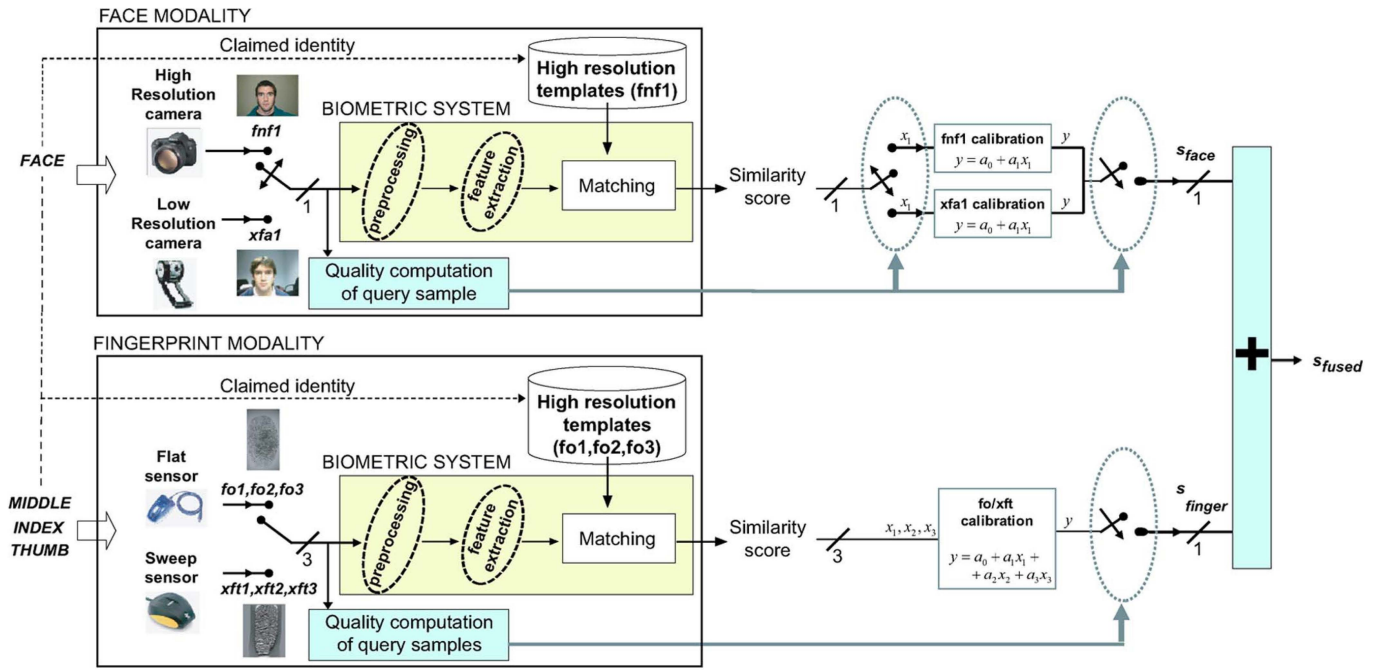

Fig. 2. Architecture of the proposed fusion strategy, including some quality-based conditional processing steps (highlighted with dashed ellipses).

Bayesian framework are considered, log-likelihood-ratios may be simply added.

The contribution of this paper is multifold. First, we summarize related works in the field of *quality-based conditional processing*. This concept is already around in the biometric community [19] but has not appeared in scientific journals yet. Second, we develop an application of this idea in the framework of the BMEC, with a novel fusion architecture that allows system interoperability by combining signals from different devices. Quality information is used to switch between different system modules depending on the data source (the sensor in our case) and to reject channels with low-quality data during the fusion. Third, while incorporation of quality measures has been done mostly by heuristically adapting the system [34], our approach easily generalizes to multiple sources of information (different modalities, matchers, acquisition devices, etc.). Newer developments and additional modalities can be easily incorporated while efficiently handling the different sources of information. The only requirement is to provide log-likelihood-ratios as output scores for the fusion. Fourth, while quality measures are often treated as scalar values, we consider them as a *vector* of measurements in this paper. Quality is intrinsically multidimensional, affected by factors of very different natures (e.g., for a face image: pose, frontalness, focus, illumination, etc.). A biometric system must adequately address this multifactor nature [21]. Other contributions are related with the experimental framework of the paper. The proposed architecture, preliminary evaluated in [10], incorporates here the quality-based score rejection step. Because of this, an additional EER improvement of about 25% is obtained. Also, the proposed system is demonstrated to outperform a set of rule-based fusion schemes used for comparison [35], [36], highlighting the effectiveness of the proposed approach.

The rest of this paper is organized as follows. In Section II, the proposed probabilistic fusion scheme is introduced and linear logistic regression fusion is described. Section III describes the evaluation framework, including the data set and protocol used in our experiments. Results obtained are presented in Section IV, and conclusions are given in Section V.

## II. PROBABILISTIC FUSION APPROACH FOR QUALITY-BASED CONDITIONAL PROCESSING

In this paper, we propose a score fusion approach that presents advantages over other methods when signals originate from heterogeneous biometric sources. We adopt a probabilistic Bayesian framework, presenting two stages. First, the similarity scores of each modality are mapped to a probabilistic log-likelihood-ratio via a procedure known as *calibration* [33]. Second, the calibrated scores, interpretable as log-likelihood-ratios, are summed. The whole process is represented in Fig. 2, and both steps are described below. Quality-based conditional processing is performed in: *1)* the normalization stage using different calibration functions depending on the device used for query acquisition, which is estimated from quality signals; and *2)* the fusion stage, discarding scores which come from low-quality sources.

### A. Proposed Approach

In order to illustrate the motivation and advantages of such approach, we first consider the verification process as a binary classification scenario, where two hypotheses (classes) are present: the *target* hypothesis ($\theta_t$: the compared biometric samples come from the same individual); and the non *target* hypothesis ($\theta_{nt}$: the compared biometric samples come from different individuals). Now, assume that the information from $M$ different matchers is the vector of scores $\mathbf{x} = [x_1, x_2, \ldots, x_M]^T$.

In a Bayesian probabilistic framework, the optimal decision is taken using the posterior probability $P(\theta_t|\mathbf{x}) = 1 - P(\theta_{nt}|\mathbf{x})$ according to the following rule:

$$\text{For a given } \mathbf{x} \begin{cases} \text{decide } \theta_t : P(\theta_t|\mathbf{x}) > P(\theta_{nt}|\mathbf{x}) \\ \text{decide } \theta_{nt} : P(\theta_t|\mathbf{x}) < P(\theta_{nt}|\mathbf{x}) \end{cases} \quad (1)$$

which can be rewritten in terms of likelihood ratio as

$$\text{For a given } \mathbf{x} \begin{cases} \text{decide } \theta_t : (p(\mathbf{x}|\theta_t)/p(\mathbf{x}|\theta_{nt})) > \tau_B \\ \text{decide } \theta_{nt} : (p(\mathbf{x}|\theta_t)/p(\mathbf{x}|\theta_{nt})) < \tau_B. \end{cases} \quad (2)$$

Here, $\tau_B$ is known as the *Bayes threshold*, and its value depends on the prior probabilities of the hypotheses $p(\theta_t)$ and $p(\theta_{nt})$ and on the decision costs. Both the priors and the costs are outside of the scope of the biometric system, but it is worth noting that if the problem is stated in terms of likelihood ratios, the optimal threshold can be set if $p(\theta_t)$ and $p(\theta_{nt})$ and the decision costs are known [32]. Therefore, the threshold is fixed and independent of the biometric system.

In order to cope with this framework, a calibration transformation $\phi(\mathbf{x})$ is applied to $\mathbf{x}$, after which the score can be interpreted as a log-likelihood-ratio [31], [33].

$$\phi(\mathbf{x}) = \log\left(p(\mathbf{x}|\theta_t)/p(\mathbf{x}|\theta_{nt})\right) = x^{cal}. \quad (3)$$

The calibration process gives *meaning* to $x^{cal}$ in the following sense. An uncalibrated score $x_i \in \mathbf{x}$ is a measure of similarity among two biometric samples; however, it is meaningless unless the distributions of target and nontarget scores are known. For instance, if $x_i = 4$, we cannot determine which hypothesis, target or nontarget, it supports the most. If we additionally know that the target distribution ranges among 2 and 4, we can then determine that $x_i$ is strongly supporting the target hypothesis. In this sense, the calibration process allows the interpretation of a single score, $x^{cal}$, as a degree of support to any of the hypotheses: if the score is higher than 0, then the support to $\theta_t$ is also higher, and vice-versa. Note that a calibrated score $x^{cal} = 0$ means no support to any hypothesis, so no discriminative information is given by such a score.

It is easy to show that if M calibrated scores $\{x_1^{cal}, x_2^{cal}, \ldots, x_M^{cal}\}$ come from statistically independent modalities, its sum also yields a calibrated fused score [32]

$$x'^{cal} = x_1^{cal} + x_2^{cal} + \cdots + x_M^{cal}. \quad (4)$$

This justifies the proposed fusion approach, where we sum calibrated scores from independent modalities. This approach presents the following advantages when dealing with signals that originate from heterogeneous biometric sources.

- The meaning of a log-likelihood-ratio is the same across different systems. This allows the comparison of biometric signals that originate from different sources (modalities, matchers, devices, etc.) in the same probabilistic range.
- The interpretability of the fused score $x'^{cal}$ as a log-likelihood-ratio allows the use of Bayes thresholds for optimal decision-making, avoiding the need of computing a new threshold each time a part of the system is changed. This is essential in operational conditions because the threshold setting critically determines the accuracy of the authentication process in many applications.

- A log-likelihood-ratio will give more support to the target or nontarget hypothesis depending on the accuracy of the matcher. When combining (summing) independent sources, the most reliable modality will have a dominant role. In other standard normalization methods [20] (as the one used in this paper as baseline, see Section III-D), all the modalities are considered to have the same weight in the fusion since scores of each modality are normalized to a similar range (e.g., [0, 1]) independently of its accuracy. This is a common problem of these normalization methods, which makes the worst modalities to yield misleading results more frequently [20].
- The sum of calibrated scores [(4)] also yields a calibrated score regardless of the number $M$ of sources. It provides an elegant and simple solution for handling a variable number of sources, allowing an infrastructure lacking homogeneity. For instance, users can be enrolled in all the biometric traits and later, depending on the access location (home, office, etc.), different modalities and/or sensors can be used. The same applies when a source is rejected due to low quality, as done in this paper.

### B. Score Calibration via Logistic Regression

Logistic regression is a well known pattern recognition technique widely used for many problems including fusion [30], [31], and more recently, calibration [33]. Its aim is to obtain an affine transformation (i.e., shifting and scaling) of an input data set in order to optimize an objective probabilistic function. For a given vector of scores $\mathbf{x} = [x_1, x_2, \ldots, x_M]^T$, the logistic regression model can be stated as follows:

$$f_{lr} = \log\frac{P(\theta_t|\mathbf{x})}{P(\theta_{nt}|\mathbf{x})} = a_0 + a_1 \cdot x_1 + \cdots + a_M \cdot x_M. \quad (5)$$

Using Bayes theorem, this expression allows the computation of the log-likelihood-ratio [32]

$$\log\frac{p(\mathbf{x}|\theta_t)}{p(\mathbf{x}|\theta_{nt})} = a_0 + a_1 \cdot x_1 + \cdots + a_M \cdot x_M - \lambda \quad (6)$$

where $\lambda$ is a factor dependent on the ratio $p(\theta_t)/p(\theta_{nt})$ [32]. However, this factor has no influence in the transformation to log-likelihood-ratios, as can be seen in [33]. In fact, it can be demonstrated [30], [31] that the optimization of the likelihood ratio in (6) can be achieved by minimizing the following objective function with respect to $\{a_0, a_1, \ldots, a_M\}$ for an arbitrary given value of the prior probabilities[1]:

$$C_{lr} = P(\theta_t)\frac{1}{N_{T_G}}\sum_{j=1}^{N_{T_G}} \log\left(1 + e^{-f_{lr}^j}\right) + P(\theta_{nt})\frac{1}{N_{T_I}}\sum_{j=1}^{N_{T_I}} \log\left(1 + e^{-f_{lr}^j}\right) \quad (7)$$

where $f_{lr}^j$ values come from applying (5) to a training set of target $(j = 1, \ldots, N_{T_G})$ and nontarget $(j = 1, \ldots, N_{T_I})$ scores for each modality $i$ ($N_{T_G}$ and $N_{T_I}$ are the number of target and nontarget scores in the training set).

---

[1] If the prior probability is not known, as it may happen in some applications, a value of 0.5 is a recommendable choice [33].

TABLE I
EXPERIMENTAL PROTOCOL

|  | | Number of matching scores per subject | |
|---|---|---|---|
| **DATASETS** | | Training (51 subjects) | Evaluation (156 subjects) |
| Session 1 | Genuine | 1 | 1 |
|  | Impostor | $103 \times 4$ | $126 \times 4$ |
|  | Purpose | Development | User adaptation |
| Session 2 | Genuine | 2 | 2 |
|  | Impostor | $103 \times 4$ | $126 \times 4$ |
|  | Purpose | Development | Test |

TABLE II
BIOMETRIC TRAITS AND BIOMETRIC DEVICES CONSIDERED

| MODE | DATA TYPE | SENSORS | CONTENTS |
|---|---|---|---|
| *fnf1* | Face still | Digital camera (high res.) | Frontal face |
| *fa1* | | Webcam (low resolution) | |
| *fo1, fo2, fo3* | Fingerprint | Optical (flat) | 1/2/3: right thumb/ |
| *ft1, ft2, ft3* | | Thermal (sweeping) | index/middle |

TABLE III
REFERENCE SYSTEMS AND QUALITY MEASURES USED

| REF. SYSTEM | QUALITY MEASURES |
|---|---|
| **FACE STILL MODALITY** | |
| | Face detection reliability, Brightness, Contrast, |
| Omniperception SDK | Focus, Bits per pixel, Spatial resolution, |
| LDA-based face verifier [38] | Illumination, Uniform Background, Background |
| | Brightness, Reflection, Glasses, Rotation in |
| | plane, Rotation in Depth, Frontalness |
| **FINGERPRINT MODALITY** | |
| NIST fingerprint system [39] | Texture richness based on local gradient [26] |

As it can be seen, the logistic regression model encourages the probabilistic interpretation of the output score in the form of a log-likelihood-ratio by means of the affine calibration transformation $\phi(\mathbf{x}) = a_0 + a_1 \cdot x_1 + \cdots + a_M \cdot x_M$ [(5)]. This process is performed for each independent modality in the system, allowing the sum fusion scheme proposed in this section. In order to perform logistic regression calibration, the freely available Focal toolkit for Matlab has been used [37].

## III. EVALUATION FRAMEWORK

### A. Data Set and Experimental Protocol

We use the set of scores of the *Access Control Scenario Evaluation* of the BioSecure Multimodal Evaluation Campaign [9] as provided to the participants. This evaluation was conducted in 2007 by the BioSecure Network of Excellence [12] as a continuation of the acquisition campaign of the Biosecure Multimodal Database [10]. The aim of this evaluation was to compare the performance of multimodal fusion algorithms, assuming that the environment is relatively well controlled and the users are supervised. We focus on the *quality-based evaluation*, whose objective was to test the capability of a fusion algorithm to cope with query biometric signals that originate from heterogeneous biometric devices.

The Biosecure Multimodal Database contains six biometric modalities [10]: face, speech, signature, fingerprint, hand, and iris. Several devices under different conditions and levels of supervision were used for the acquisition. In this paper, we use a subset of 333 persons designed for the purpose of the *Access Control Evaluation*. This subset was collected over two sessions, separated by about one month interval, with two biometric samples per device and session. The first sample of session one was considered as the template, whereas the remaining three samples were considered as query data.

Among the 333 subjects, 207 were considered "clients" for whom a template was created: 51 "clients" for *training* (whose scores and identity labels were provided to the participants to tune their algorithms) and 156 for *evaluation* (whose scores, mixed genuine and impostor claims, were provided to the participants to be fused without identity labels, which were sequestered by the evaluation organizers to evaluate the competing algorithms). The remaining 126 subjects served as "zero-effort impostors" (i.e., no template is created for them). The experimental protocol is summarized in Table I. The *training* impostor set of scores of Session 1 contains $103 \times 4$ samples per subject, which means that when the reference subject is considered a template, all the 4 samples of the half of the remaining 206 subjects are considered impostors. The other half are used as impostors in Session 2. This ensures that the impostors used in Sessions 1 and 2 of the training set are not the same. Note that the *evaluation* impostor score sets contain the 126 zero-effort impostors, so a fusion algorithm will not have already "seen" the impostors during its training stage, avoiding systematic and optimistic bias of performance. Prior to the evaluation, the *training* set of scores was released to the participants to tune their algorithms. It was recommended to use only Session 2 as training data since Session 1 may be optimistically biased due to the use of template and query data acquired on the same session. In this paper, we follow this recommendation, using only Session 2 of the training set for training our algorithms. Session 1 of the evaluation set is intended for user-adapted fusion [8], whereas Session 2 is for testing purposes. The work reported here is not user-adaptive and therefore, it will only be run on Session 2 of the evaluation set. The division of training and testing and query and enrolment in the data was the same for all the participants, which is also the same used in the experiments of this paper.

### B. Face and Fingerprint Systems

The *Access Control Evaluation* only considered face and fingerprint modalities [3] (see Table II). Several reference systems and quality measures were used. These reference systems are summarized in Table III. Low- and high-quality still frontal face images were collected with two different cameras (denoted as *fa1* and *fnf1*, respectively). The system used is an LDA-based face verifier [38], and the 14 face quality measures indicated in Table III were computed using the proprietary Omniperception SDK.[2] The fingerprint data was collected with an optical and a thermal sensor, denoted as $fo\{n\}$ and $ft\{n\}$, respectively, with $n = \{1 = \text{thumb}, 2 = \text{index}, 3 = \text{middle}\}$ fingers of the right hand. The system used is the NIST fingerprint system [39], whereas the quality measure is based on averaging local gradients [26]. In Fig. 3, the biometric sensors as well as acquisition samples of the modalities used in the evaluation are shown.

---
[2]http://www.omniperception.com.

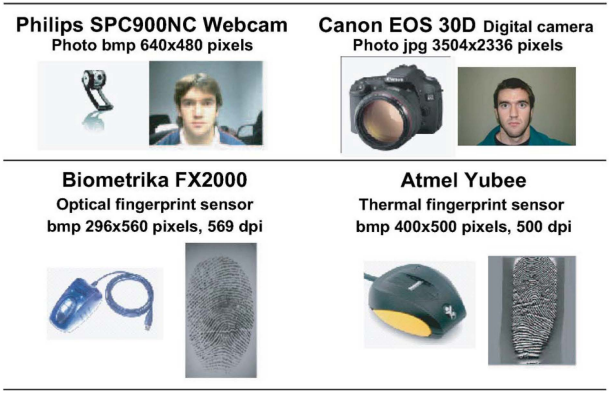

Fig. 3. Hardware devices and samples of the modalities considered in the *Access Control Evaluation*. Top row: face modality (left/right: low/high resolution camera). Bottom row: fingerprint modality (left: optical sensor with flat positioning of the finger; right: thermal sensor with finger sweeping).

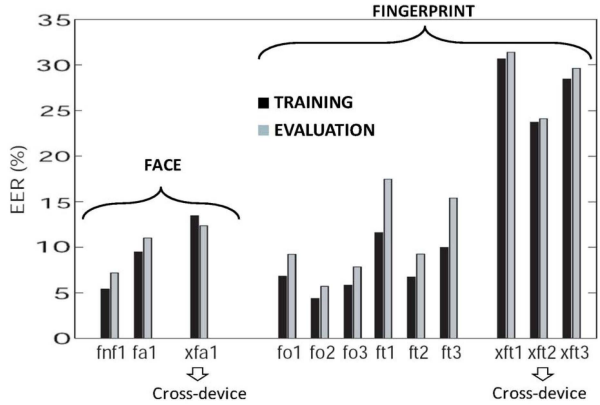

Fig. 4. Performance of the modalities of Table II in the training and evaluation sets defined in Table I. Results of cross-device matching are also shown.

### C. Cross-Device Evaluation

In the *quality-based evaluation*, matching can occur between a biometric template acquired using one device and a query biometric data acquired using another device. Template data is always acquired with a high-quality device [i.e., giving better verification performance (see Fig. 4)] and query data may be acquired using a high- or a low-quality device. This is a reasonable operational assumption (i.e., quality control is applied at least during enrolment), although this may not be true in some unsupervised environments [21]. The channels of data considered are face and the three right fingerprints, denoted as *fnf1*, *fo1*, *fo2*, and *fo3* (see Table II). In case of cross device matching, these channels are denoted as *xfa1*, *xft1*, *xft2*, and *xft3*. The development set distributed to the participants consisted of scores and quality measures of all 8 channels. The (sequestered) evaluation set, on the other hand, contained only 4 channels of data as a result of mixing *fnf1/xfa1* and *fo{n}/ft{n}*, with $n = \{1 = \text{thumb}, 2 = \text{index}, 3 = \text{middle}\}$. These four channels of data can be any of the combinations of Table IV (for a given access, all fingerprints were acquired with the same device). The performance of the different modalities on the training and evaluation sets are shown in Fig. 4.

It should be noted that there was missing data in the sets of scores because some matchings or quality estimates could not be computed by the algorithms used in the evaluation. In the training set, one missing score in the *xft1* channel and 51 scores in the *fo3* channel are found (less than 1%). In the *test* set, about 3% of the scores in the fingerprint channels and 16% in the face channel are missing. It is not the aim of this paper to deal with missing data in multibiometrics, so prior to the experiments, we have corrected the missing values of the *training* set as follows. When a genuine (impostor) score of a specific sensor is missing, its value is set to the mean value of the remaining valid genuine (impostor) scores. Similarly, when a quality measure is missing, its value is set to the mean value of the remaining valid measures. For the *evaluation* set, it is not known in advance if we are dealing with a genuine or an impostor access, so we use a different strategy. When a fingerprint score of an access is missing, its value is set to the mean value of the remaining valid scores prior to the fusion (the same applies to the quality values). If an entire modality is missing, it is not used in the fusion. If both modalities are missing, the fused score is set to the threshold value at the EER point on the *training* set. This was the procedure followed in our submission to the *quality-based evaluation* of Biosecure, where the rejection of an access was not allowed [9]. To be consistent with the evaluation, this procedure is also used in the experiments reported in this paper, unless indicated.

### D. Baseline Fusion Scheme

To compare the performance of the proposed probabilistic fusion approach (Section II), we use a set of baseline rule-based fusion procedures over normalized scores based on the arithmetic mean, the minimum and the maximum [36]. These schemes have been widely used to combine multiple classifiers in biometric authentication with good results [35], [36]. The use of these fusion rules is motivated by their simplicity, as complex fusion approaches need training and their superiority over fixed fusion approaches cannot even be guaranteed (e.g., see [40]). Matching scores are first normalized to be similarity scores in the [0, 1] range using tanh-estimators [20]

$$s' = 0.5 \{\tanh (0.01 ((s - \mu_s)/\sigma_s)) + 1\} \qquad (8)$$

where $s$ is the raw similarity score, $s'$ is the normalized similarity score, and $\mu_s$ and $\sigma_s$ are the estimated mean and standard deviation of the genuine score distribution, respectively. The tanh-estimator is used because it is referred to as a robust and highly efficient normalization method [20]. Similar to the architecture proposed in Fig. 2, face and fingerprint scores are normalized separately and subsequently fused with the mentioned rules.

| # | MODALITIES | FACE | FINGERPR. | DEVICE MISMATCH |
|---|---|---|---|---|
| 1 | (*fnf1/fo1/fo2/fo3*) | HR | FA | no mismatch |
| 2 | (*fnf1/xft1/xft2/xft3*) | HR | SA | fingerprint sensor |
| 3 | (*xfa1/fo1/fo2/fo3*) | LR | FA | face sensor |
| 4 | (*xfa1/xft1/xft2/xft3*) | LR | SA | fingerprint+face sensors |

## IV. EXPERIMENTS

The experiments are divided into four parts.

1) *Estimation of the Acquisition Device From Quality Measures.* There are four subproblems that can be identified in addressing cross-device matching depending on the following assumptions [29]: *a)* whether a set of possible devices is known or not; and *b)* whether the actual acquisition device is known or not. Here, we will address the case where the set of devices is known, but further distinguish the subproblems where the actual device used in the acquisition is known or not. When the actual device is not known, we will show that it can be inferred using quality measures from the raw biometric data.

2) *System Performance With Quality-Based Device Estimation.* In this experiment, the proposed fusion approach is compared to the baseline fusion rules using device estimation based on quality measures. For comparison purposes, we also show the performance when the actual acquisition device is known.

3) *Sensor Interoperability Analysis.* Here, we test the capabilities of the proposed fusion approach to cope with biometric data from heterogeneous devices. We report the performance of the different combinations of channels of data described in Section III-C/Table IV, both for the proposed fusion approach and for the best baseline fusion rule according to the results of Part 2.

4) *System Performance With Quality-Based Score Rejection.* This experiment is aimed to test the effects of rejecting channels with low quality data during the fusion. According to our experiments, an EER improvement of 25% is achieved by incorporating a quality-based score rejection scheme in the proposed fusion approach.

### A. Estimation of the Acquisition Device From Quality Measures

According to the protocol of the *quality-based evaluation* [9], no information was given regarding the device used for query acquisition. In this scenario, we were interested in exploring the potential benefits of conditional processing based on a prediction of the input device. For this purpose, we used the quality measures provided assuming the following.

- If the template and the query were from the same device (i.e., *fnf1, fo1, fo2, fo3*), both images would have similar quality values and they would be high.
- If the template and the query were from different devices (i.e., *xfa1, xft1, xft2, xft3*), the quality value of the template would be higher than the quality value of the query, and the quality value of the query would be low.

To estimate the device, we used a quadratic discriminant function with multivariate normal densities for each class [32]. For the face modality, we used the 14 quality measures of the query image (see Table III). For the fingerprint modality, we derived the following 8 parameters from the quality of the templates ($Q_{ti}$) and queries ($Q_{qi}$) of the three scores corresponding to each access ($i = 1, 2, 3$): 1) Number of fingerprint scores such as $Q_{ti} > Q_{qi}$; 2) $\max(Q_{qi})$; 3) $\max(|Q_{ti} - Q_{qi}|)$; 4) $\min(Q_{qi})$; 5) $\min(|Q_{ti} - Q_{qi}|)$; 6) $\text{mean}(Q_{qi})$; 7) $\text{mean}(|Q_{ti} - Q_{qi}|)$; and 8) $\max(Q_{ti} - Q_{qi})$.

TABLE V
QUALITY FEATURE COMBINATION FOR THE ESTIMATION OF THE DEVICE USED FOR THE QUERY ACQUISITION. RESULTS SHOW THE ERROR RATES IN THE ESTIMATION OF THE DIFFERENT DEVICES

| | TRAINING SET | | EVALUATION SET | |
|---|---|---|---|---|
| **Face feature** | **Error fnf1** | **Error xfa1** | **Error fnf1** | **Error xfa1** |
| 8 | 0.04% | 0.37% | 4.43% | 17.59% |
| 6-8 | 0.04% | 0.37% | 4.14% | 17.79% |
| 8-9 | 0.04% | 0.37% | 4.14% | 19.44% |
| 6-8-9 | 0.04% | 0.12% | 4.14% | 18.82% |
| **Fingerprint feature** | **Error fo** | **Error xft** | **Error fo** | **Error xft** |
| 2 | 21.81% | 8.03% | 29.26% | 10.73% |
| 1-2 | 22.89% | 10.47% | 28.86% | 14.53% |
| 2-3-6 | 22.08% | 9.41% | 21.69% | 12.82% |

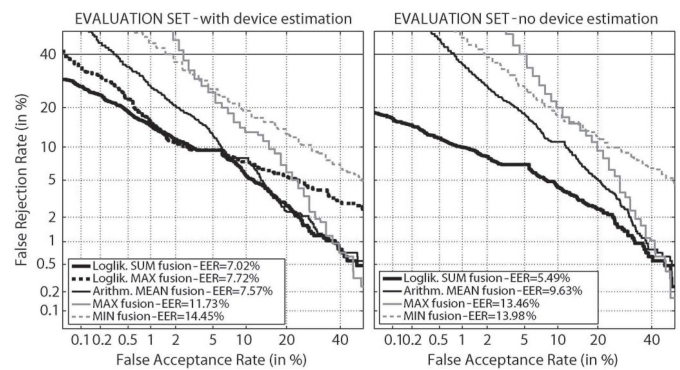

Fig. 5. Verification results of the proposed log-likelihood fusion (Loglik.) together with baseline fusion rules used for comparison (Loglik. SUM is further studied in the present paper, Loglik. MAX was the approach submitted by the authors to the quality-based Biosecure Evaluation). (Left plot) With device estimation using quality measures. (Right plot) Without device estimation (knowing the actual device used in each access).

We tested all the combinations of 1, 2, and 3 quality features in order to determine the device used for the query acquisition. Results of the best cases are shown in Table V. For the face modality, a remarkably low error rate is obtained using the training set, even with only one parameter. This is not true for the evaluation set, which could be due to the small size of the data set provided for training ($51 \times 103 \times 4 = 21\,012$ impostor scores but only $51 \times 2 = 102$ genuine scores, see Table I). On the other hand, we observe high error rates in the estimation for the fingerprint modality in both data sets. Interestingly enough, the estimation fails mostly with the optical sensor.

### B. System Performance With Quality-Based Device Estimation

Based on the results of the input device estimation on the training set, we trained a score normalization function for each face modality (*fnf1, xfa1*), and a unique fusion function for both fingerprint modalities (*fo, xft*), as shown in Fig. 2. This was our approach submitted to the *quality-based evaluation*, but using the MAX rule of the two calibrated scores instead of the SUM rule of this paper [10]. Results of both approaches are shown in Fig. 5(left), together with the baseline fusion rules used for comparison. As observed in Table V, the device estimation did not perform well on the evaluation set (a fact not known until the evaluation set was released after the evaluation). Therefore, with the aim of evaluating the effects of such a bad device estimation, we also depict in Fig. 5(right) the results considering the actual device used in each access, training in this case a modality-specific score normalization function. Although that was not the protocol in the evaluation, it is reasonable to assume that the specific sensor used in an operational environment is known.

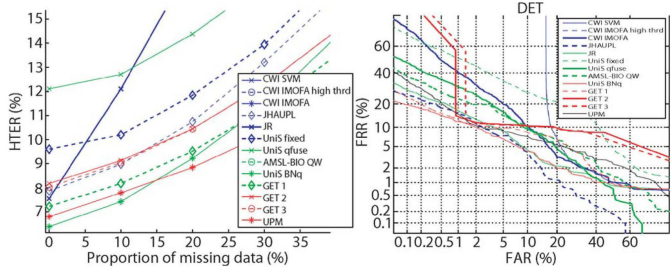

Fig. 6. Results of the *quality-based evaluation* of the BioSecure Multimodal Evaluation Campaign on the evaluation set (extracted from [9]). Our approach is marked as "UPM." (Left) Fusion system performance versus proportion of missing data. (Right) Fusion system performance in DET.

As can be observed in Fig. 5(left), the proposed approach based on log-likelihood SUM fusion results in the best performance, outperforming all the baseline fusion rules (including the log-likelihood MAX fusion). Only the arithmetic mean and the maximum rules result in similar performance for low FRR values, being this difference higher when knowing the actual device used in each access [Fig. 5(right)]. Finally, comparing Fig. 5(right) to Fig. 5(left), we can also see the decrease in performance when using device estimation in comparison to knowing the actual device used in each access due to the bad device estimation obtained in the evaluation set (from 5.49% to 7.02% in the EER).

We plot in Fig. 6 the results of the BMEC (extracted from [9]). Two indicators were used for performance assessment: equal error rate and half total error rate. EER is defined as the operating point where FAR = FRR. There is only one unique threshold satisfying this condition. HTER is the average of FAR and FRR at a particular threshold supplied by the participants (in our case, the threshold is equal to zero, meaning no support to any of the target or nontarget hypotheses). Hence, although a fusion system may have a low EER, its HTER can still be relatively high [9] due to a badly estimated threshold. As can be observed in Fig. 6, the log-likelihood MAX fusion submitted to the evaluation was ranked second out of 13 participants in terms of HTER and fourth in terms of EER. We also provide in Table VI a brief description of the top performing systems. The algorithms submitted were of very diverse nature [9]: *1) generative classifiers*, in which class-dependent densities are first estimated and decisions are taken using Bayesian classifiers (e.g., GET1 and UniS BNq algorithms), Bayesian belief networks [41] (e.g., JHUAPL algorithm), or the Dempster–Shafer theory of evidence [42] (e.g., JR algorithm); *2) discriminative classifiers*, where the decision boundary is directly estimated using SVMs, logistic regression (e.g., our algorithm, marked as "UPM"), etc.; and *3) transformation-based*, which first transforms the scores of each biometric system into a comparable range, e.g., [0,1], and then combine the normalized scores using a fixed rule such as the sum or the product [36]. The latter is the strategy followed by the baseline fusion scheme used in this paper (Section III-D).

TABLE VI
BRIEF DESCRIPTION OF TOP PERFORMING SYSTEMS AT THE
*QUALITY-BASED EVALUATION* OF THE BIOSECURE MULTIMODAL EVALUATION
CAMPAIGN (MORE DETAILS CAN BE FOUND IN [9])

| Name | Characteristics |
|---|---|
| JHUAPL | Generative classifier using a bayesian belief network [41]. Quality data is clustered by binning it to compute an histogram. |
| JR | Generative classifier using Dempster-Shafer of evidence [42] |
| GET1 | Generative classifier using a bayesian classifier. Estimation of densities with Gaussian Mixture Models [43]. Fusion of normalized scores with the average rule. Quality information is not used. |
| UniS BNq | Generative classifier using a bayesian classifier [29]. Fusion of normalized scores using the sum rule. Quality data is clustered by using Gaussian Mixture Models [43]. |
| UPM | Discriminative classifier using linear logistic regression, mapping scores into log-likelihood ratios. Fusion of normalized scores using the max rule. The mapping function is chosen according to the inferred device, which is done using the quality information. |

TABLE VII
VERIFICATION RESULTS OF THE FUSION IN TERMS OF EER (%) FOR THE
FOUR DIFFERENT MIXTURES DEFINED IN TABLE IV ON THE EVALUATION
SET. THE RELATIVE EER INCREASE WITH RESPECT TO THE BEST
MODALITY INVOLVED (SEE FIG. 4) IS ALSO GIVEN IN BRACKETS

| Mixture | Modalities | With device estimation | | With correct device model | |
|---|---|---|---|---|---|
| | | loglik SUM | Arith. mean | loglik SUM | Arith. mean |
| 1 | (fnf1/fo1/ fo2/fo3) | 2.88% (-50.30%) | 1.64% (-71.70%) | 1.55% (-73.25%) | 1.61% (-72.22%) |
| 2 | (fnf1/xft1/ xft2/xft3) | 6.69% (-16.49%) | 11.18% (+39.56%) | 6.97% (-12.99%) | 12.38% (+54.54%) |
| 3 | (xfa1/fo1/ fo2/fo3) | 2.55% (-56.00%) | 2.26% (-61.00%) | 1.91% (-67.04%) | 1.62% (-72.05%) |
| 4 | (xfa1/xft1/ xft2/xft3) | 9.24% (-29.81%) | 11.35% (-13.79%) | 9.26% (-29.66%) | 12.05% (-8.47%) |
| | **ALL** | **7.02%** | **7.57%** | **5.49%** | **9.63%** |

According to Fig. 6(left), the top two systems are UniS BNq and our system, "UPM." These two systems are device-specific: they first estimate how probable the channel of data is from the observed quality measures, and then use the corresponding device-dependent fusion function. The next best system is GET1, which is a bayesian classifier whose class-conditional densities are estimated using GMMs [43]. This system does not use quality measures, so it does not change its fusion strategy under cross-device matching. Some of the systems that use quality measures are not among the best systems in terms of HTER because they did not use the right threshold, e.g., JHUAPL. The performance can be assessed in the DET curve [Fig. 6(right)] independent of the decision threshold. Here, we can observe the good performance of JHUAPL. It can be also observed that JHUAPL dominates for low FRRs, whereas UniS BNq performs very well for low FARs. We also can see that the top systems have about 6.0%–6.5% EER, with our system having an EER of 7.72%. Most of the participating systems are below 10% EER. It is worth noting that one baseline fusion rule used in the experiments of this paper works better than some of the participating systems, which are based on more complex trained approaches [an EER of 7.57% is obtained with the arithmetic mean rule; see Fig. 5(left)]. A weakness observed in our system is its degradation at low FRR values with respect to other systems having worse EER [see Fig. 6(right)]. This is improved with the log-likelihood SUM fusion used in this paper [Fig. 5(left)], also observing an improvement at low FAR values.

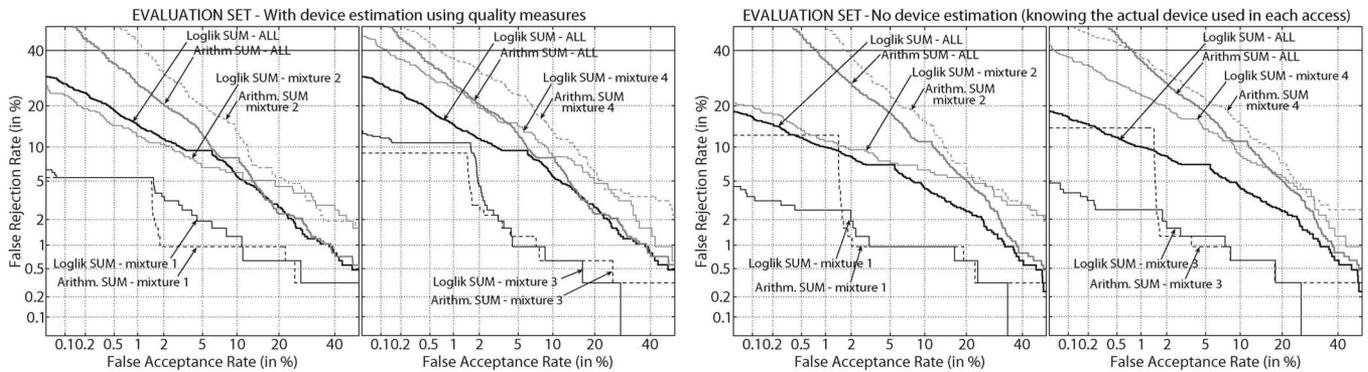

Fig. 7.   Verification results of the fusion for the different mixtures defined in Table IV. (Left plots) With device estimation using quality measures. (Right plots) Without device estimation (knowing the actual device used in each access).

### C.  Sensor Interoperability Analysis

To evaluate the capability to cope with query biometric signals from heterogeneous devices, we report in Table VII the performance of the four possible combinations for an access using the proposed log-likelihood sum fusion rule $s_{fused}$ and the best baseline fusion rule (the arithmetic mean). DET curves are also plotted in Fig. 7. It can be observed that for the mixtures involving only the optical sensor (1 and 3), there are no big differences in performance between the two fusion schemes. On the other hand, for the mixtures involving mismatched fingerprint devices (2 and 4), the proposed fusion scheme outperforms the baseline fusion rule. This is specially evident for mixture 2, which does not involve mismatched face devices (only the high resolution camera). We can also see that the proposed scheme performs best in overall terms, i.e., when pooling all the mixtures.

It is also worth noting that the best mixtures (1 and 3) do not use mismatched fingerprint devices and they also result in the highest relative improvement with respect to the best individual modality involved, as observed in Table VII. The mixture involving both mismatched fingerprint and face devices (4) always performs the worst. However, about a 30% of improvement is obtained in terms of EER for this mixture when fusing, as compared to the best single modality.

### D.  System Performance With Quality-Based Score Rejection

An operational approach to incorporate quality information in biometric systems is to reject low-quality samples, as proposed in several studies [14], [21]. But this can be inconvenient to users who are asked to be recaptured when particular samples are of low quality, or even to make a biometric system unsuitable to individuals whose data is not consistently of enough quality. This can be overcome by multibiometric systems [2], allowing the use of alternative sources of biometric information.

Here, we have tested this quality-based modality rejection by not considering in the fusion scores having a quality value lower than a specific threshold. The quality value of a matching score is defined as $\min(Q_t, Q_q)$, where $Q_t$, $Q_q$ are the qualities of the template and query biometric samples, respectively, corresponding to the matching. Thus, the worse of the two biometric samples drives the score [21]. For a given access (consisting of a face sample and three fingerprints; see Fig. 2), fingerprint scores with quality lower than the threshold are replaced with the fingerprint score having the maximum quality value. If the three fingerprint scores have their quality lower than the threshold, then the fingerprint modality is entirely discarded. In order to set the "optimum" quality threshold, we used the verification performance of the fingerprint and face modalities as scores with the lowest quality value are discarded, as shown in Fig. 8. Thresholds are set for each modality by choosing the value that minimizes the EER on the training set (indicated in Fig. 8 as vertical lines). Except for the *xft* fingerprint modality (template and query with flat and sweep acquisition, respectively), an EER reduction is also observed on the evaluation set for the selected thresholds.

Once the optimum thresholds are selected, we evaluate the performance of the proposed log-likelihood sum fusion on the mixtures defined in Table I by separately discarding face or fingerprint scores. Results are shown in Table VIII. In all cases, an EER decrease (or at least, no significant increase) is observed except when discarding scores of the *xft* modality. This is consistent with the results reported on Fig. 8, where no reduction on the EER was observed on the evaluation set for the selected quality threshold. Based on these results, no threshold will be subsequently applied to the *xft* modality.

Finally, we jointly apply the score quality-based rejection in all the modalities using the optimum thresholds selected. To be consistent with the constraints of the BMEC [9], where no access can be rejected, the resulting fused score is set to 0 if all the quality measures of an access are lower than the thresholds. In the proposed log-likelihood fusion strategy, this means that there is the same likelihood if signals are assumed to be originated or not by the given subject. This is an advantage of our probabilistic approach, as no value needs to be tuned as output when no information about modalities is given. However, we also report results discarding these accesses of the computation of the error rates to show the benefits of this policy. In Fig. 9, we show the number of accesses per modality that do not comply with the quality requirements, showing that the fusion allows the recovery of a significant number of the rejected accesses. Verification results of the fusion with the proposed quality-based rejection scheme are shown in Table IX and Fig. 10.

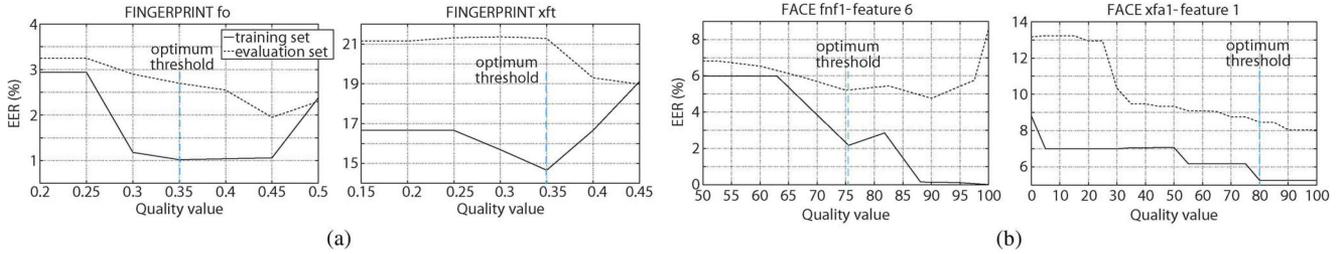

Fig. 8. Verification performance in terms of EER for the fingerprint and face modalities as scores with the lowest quality value are discarded. The quality value of a matching score is defined as $\min(Q_t, Q_q)$, where $Q_t$ and $Q_q$ are the qualities of the template and query biometric samples, respectively, corresponding to the matching. For the fingerprint modality, the quality feature of Table III is used for $Q_t$ and $Q_q$. Due to space constraints, results for the face modality are shown using only the quality feature of Table III that results in the highest improvement of the EER. (a) Fingerprint. (Left) Template and query with flat acquisition (*fo*). (Right) Template and query with flat and sweep acquisition, respectively (*xft*). (b) Face. (Left) Template and query with a high-resolution camera (*fnf1*). (Right) Template and query with high- and low-resolution cameras, respectively (*xfa1*).

TABLE VIII

PERFORMANCE (IN EER) OF THE PROPOSED LOG-LIKELIHOOD SUM FUSION ON THE MIXTURES OF TABLE IV AS SCORES WITH QUALITY VALUE LOWER THAN A PREDEFINED THRESHOLD ARE DISCARDED FROM THE FUSION. RESULTS ARE SHOWN BY EITHER DISCARDING FACE OR FINGERPRINT SCORES, TOGETHER WITH THE RESULTING RELATIVE EER INCREASE IN BRACKETS (REFERENCE RESULTS WITHOUT QUALITY-BASED SCORE REJECTION ARE SHOWN IN TABLE VII, FIFTH COLUMN). THRESHOLD VALUES ARE SELECTED ON THE BASIS OF FIG. 8

| Modality discarded | EVALUATION SET mixture | | | | ALL |
|---|---|---|---|---|---|
| | 1 | 2 | 3 | 4 | |
| Face *fnf1* | 1.59% (+2.58%) | 5.73% (-17.79%) | - | - | 5.41% (-1.45%) |
| Face *xfa1* | - | - | 1.33% (-30.37%) | 8.29% (-10.48%) | 5.12% (-6.74%) |
| Fingerprint *fo* | 1.51% (-2.58%) | - | 1.40% (-26.70%) | - | 5.57% (+1.46%) |
| Fingerprint *xft* | - | 6.77% (-2.87%) | - | 11.15% (+20.41%) | 5.91% (+7.65%) |

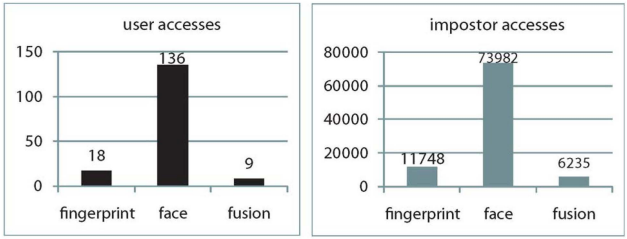

Fig. 9. Incorporation of quality information in the fusion stage. Results show the number of accesses per modality with quality value lower than the predefined thresholds. It can be observed that the fusion reduces significantly the number of rejected accesses.

TABLE IX

VERIFICATION RESULTS OF THE FUSION ON THE MIXTURES DEFINED IN TABLE II IN TERMS OF EER (%) FOR THE EVALUATION SET INCORPORATING QUALITY INFORMATION IN THE FUSION STAGE (WITHOUT DEVICE ESTIMATION). THE RELATIVE EER INCREASE AS A RESULT OF QUALITY INCORPORATION IS ALSO SHOWN (IN BRACKETS)

| mixt. | Modalities | No quality | Quality | Quality+ rejection |
|---|---|---|---|---|
| 1 | (fnf1/fo1/fo2/fo3) | 1.55% | 1.56% (-0.65%) | 1.28% (-17.42%) |
| 2 | (fnf1/xft1/xft2/xft3) | 6.97% | 5.73% (-17.79%) | 5.45% (-21.81%) |
| 3 | (xfa1/fo1/fo2/fo3) | 1.91% | 1.30% (-31.94%) | 0.96% (-49.74%) |
| 4 | (xfa1/xft1/xft2/xft3) | 9.26% | 8.29% (-10.48%) | 7.48% (-19.22%) |
| | **ALL** | 5.49% | 4.45% (-18.94%) | 4.17% (-24.04%) |

It is remarkable that even when keeping invalid accesses in the fusion, a performance improvement is obtained (see Fig. 10, curve "quality"). An additional improvement results from dis-carding these accesses (curve "quality and rejection"). It is also observed from Table IX that the highest improvement is obtained for the mixture incorporating quality-based rejection both on the fingerprint and face modalities (mixture 3). Worth noting, too, is that the mixture involving both mismatched face and fingerprint devices (mixture 4) also results in a considerable improvement. The mixture having the smallest improvement is the one involving no mismatched devices (mixture 1).

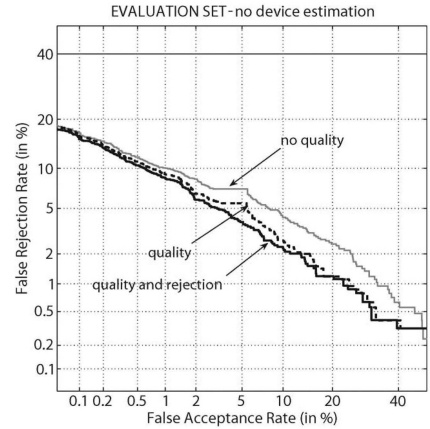

Fig. 10. Verification results of the proposed fusion incorporating quality information in the fusion stage (without device estimation).

## V. CONCLUSION

As biometric technology is increasingly deployed, it will be a common situation to replace parts of operational systems with newer designs and/or to operate with information from different sources [3]. The recent *quality-based evaluation* of the BioSecure Multimodal Evaluation Campaign [9] was aimed to compare the performance of different multimodal biometric fusion architectures and algorithms when biometric signals originate from heterogeneous devices in mismatched conditions. This evaluation operated at the matching score level, providing participants with different sets of scores which were obtained using several reference systems. Quality information of the associated biometric signals was also provided with the aim of adapting the fusion algorithms to the different devices, a strategy which is referred to as *quality-based conditional processing* [19].

In this paper, we have described the ATVS-UAM fusion strategy submitted to this evaluation [10], with very good results (second out of 13 participants in terms of HTER and fourth in terms of EER [9]). In our approach, output scores of the individual matchers are first mapped to log-likelihood-ratios by linear logistic regression prior to the fusion stage. The proposed strategy allows the efficient combination of scores that originate from different biometric sources (modalities, matchers, devices, etc.) since they are in a comparable probabilistic domain, and they are generalizable to newer developments or additional modalities. Quality-based conditional processing is carried out in two stages of the proposed strategy: by estimating the device used in each access in order to switch between different linear logistic regression modules and by rejecting scores from low-quality biometric samples. Worth noting too, is that a considerable performance improvement has been obtained when applying the quality-based score rejection. Although entirely rejecting low-quality data may be suboptimal, it constitutes a starting point on how to use low-quality data in a multibiometric system. Although low-quality data may not be able to drive as strong an inference as high-quality data, its inclusion in the fusion will be the source of future work. Quality analysis is also of crucial interest in new challenging scenarios as a result of noncooperative and/or at a distance environments [44], [45].

The proposed fusion approach is also able to cope easily with missing values of any modality. In the Biosecure *quality-based evaluation*, the robustness of the submitted algorithms against missing values was also evaluated. The proposed fusion scheme also obtained remarkable results [9], being the first in terms of HTER when the proportion of missing data was increased (i.e., higher than 20%; see Fig. 6). This encourages us to further study how to handle missing biometric data in multi-biometric scenarios.

## Acknowledgment

The primary author thanks Consejeria de Educacion de la Comunidad de Madrid and Fondo Social Europeo for supporting his Ph.D. studies.

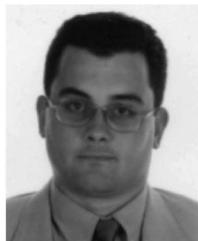

**Fernando Alonso-Fernandez** (M'09) received the M.S. and Ph.D. degrees in electrical engineering from the Universidad Politecnica de Madrid, Madrid, Spain, in 2003 and 2008, respectively.

Since 2004, he has been with the Biometric Recognition Group—ATVS, Universidad Autonoma de Madrid, Madrid, Spain, where he currently holds a Juan de la Cierva Postdoctoral Fellowship. He is actively involved in European projects focused on biometrics (e.g., FP6 Biosecure NoE, COST 2101).

He participated in the development of the ATVS systems for the BioSecure Multimodal Evaluation Campaign 2007, the Signature Competition SigComp 2009, and the Fingerprint Liveness Detection Competition 2009. His research interests include signal and image processing and pattern recognition and biometrics. He has over 30 international contributions at refereed conferences and journals and has authored several book chapters.

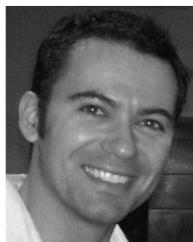

**Julian Fierrez** (M'07) received the M.S. and Ph.D. degrees in telecommunications engineering from the Universidad Politecnica de Madrid (UPM), Madrid, Spain, in 2001 and 2006, respectively.

Since 2002, he has been with the Biometric Recognition Group—ATVS, first at UPM, and since 2004, at the Universidad Autonoma de Madrid, where he currently holds a Marie Curie Postdoctoral Fellowship. As part of that fellowship, he has spent two years as a Visiting Researcher at Michigan State University, East Lansing. He is actively involved in European projects focused on biometrics (e.g., FP6 BioSec IP, FP6 BioSecure NoE, FP7 BBFor2 Marie Curie ITN). His research interests include signal and image processing, pattern recognition and biometrics with emphasis on signature and fingerprint verification, multi-biometrics, biometric databases, and system security.

Dr. Fierrez is the recipient of a number of research distinctions, including best Ph.D. thesis in computer vision and pattern recognition in 2005–2007 by the IAPR Spanish liaison (AERFAI), Motorola best student paper at ICB 2006, EBF European Biometric Industry Award 2006, and IBM best student paper at ICPR 2008.

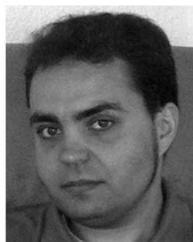

**Daniel Ramos** (M'08) received the M.S. degree in electrical engineering and the Ph.D. degree in telecommunication engineering from the Universidad Autonoma de Madrid, Madrid, Spain, in 2001 and 2007, respectively.

Since 2003, he has been with the Biometric Recognition Group—ATVS, and since 2006, he has been an Assistant Professor at the Universidad Autonoma de Madrid. He has participated in the development of the ATVS Speaker and language recognition systems since 2004. He has been part of several organizing and scientific committees in the field.

Dr. Ramos has been the recipient of several awards and distinctions such as the IBM Research Best Student Paper Award at Odyssey 2006.

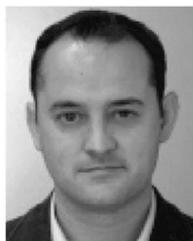

**Joaquin Gonzalez-Rodriguez** (M'96) received the M.S. and Ph.D. degrees (*cum laude*) in electrical engineering from the Universidad Politecnica de Madrid (UPM), Madird, Spain, in 1994 and 1999, respectively.

He is Founder and Co-director of the Biometric Recognition Group—ATVS. He is currently an Associate Professor with the Universidad Autonoma de Madrid, Madrid, Spain, where he leads the Speech Group of ATVS. He led the ATVS participation in the NIST Speaker (2001, 2002, 2004, 2005, 2006, and 2008) and Language Recognition Evaluation (2005 and 2007) and the NFI-TNO Forensic Speaker Recognition Evaluation (2003).

Dr. Gonzalez-Rodriguez is a member of the International Speech Communication Association (ISCA) and IEEE Signal Processing Society. Since 2000, he has been an invited member of the Forensic Speech and Audio Analysis Working Group in the European Network of Forensic Science Institutes. He is also a member of the Program Committee of the ISCA Odyssey conferences on Speaker and Language Recognition and was the Vice-Chair of Odyssey 2004 in Toledo (Spain). He was awarded the Google Research Award in 2009.